\documentclass[usenatbib]{mn2e}

\usepackage{epsfig}

\begin{document}

\title[The Impact of Oscillations on Planet Searches]{The Impact of Stellar
  Oscillations on Doppler Velocity Planet Searches}
\author[S.~J.~O'Toole et al.]{S.~J.~O'Toole,$^1$ C.~G.~Tinney,$^{1,2}$
  H.~R.~A.~Jones$^3$ \\
$^1$Anglo-Australian Observatory, PO Box 296, Epping 1710, Australia \\
$^2$Department of Astrophysics, School of Physics, University of NSW,
  2052, Australia \\
$^3$Centre for Astrophysical Research, University of Hertfordshire,
Hatfield, AL\,10\,9AB, UK}

\maketitle
\begin{abstract}
We present a quantitative investigation of the effect of stellar
oscillations on Doppler velocity planet searches. Using data from four
asteroseismological observation campaigns, we find a power law
relationship between the noise impact of these oscillations on Doppler
velocities and both the luminosity-to-mass of the target stars, and
observed integration times.
Including the impact of oscillation jitter should improve the quality of
Keplerian fits to Doppler velocity data. The scale of the effect these
oscillations have on Doppler velocity measurements is smaller than
that produced by stellar activity, but is most significant for giant
and subgiant stars, and at short integration times (i.e. less than a few
minutes). Such short observation times tend to be used only for very
bright stars. However, since it is these very same stars that tend to
be targeted for the highest precision observations, as planet searches
probe to lower and lower planet masses, oscillation noise for these
stars can be significant and needs to be accounted for in observing
strategies.
\end{abstract}

\section{Introduction}
\label{sec:intro}

Uncertainties in high-precision Doppler velocity measurements are now
reaching $\sim$\,1\,m\,s$^{-1}$ on a regular basis
\citep[e.g.][]{OBT07,PCM07}. The science drivers behind the quest for
ever-greater precision are the detection of Earth-mass planets in
short-period orbits and Solar System analogues, as well as very low
amplitude stellar oscillations. Understanding both increasingly subtle
instrumental variations, and intrinsic stellar variability, is now more
important than ever. 

The term jitter has been coined
to describe the noise imposed on precision radial velocity
programs by a star's intrinsic instability and has been investigated
by \citet{SD97} and \citet{SBM98}, with the effect particularly
detrimental on giant stars. Until now, only the
stellar activity component of jitter has been
investigated quantitatively. \citet{Wright05} examined the
jitter of targets in the Lick and Keck Planet Searches and derived an
empirical relationship based on colour, activity and
absolute magnitude. This has allowed the inclusion of an
additional term to the measurement uncertainties used when fitting
Keplerians to velocity data, and therefore better modelling of the
scatter about these fits.

The dominant oscillations in the solar-like stars are nonradial
$p$-modes stochastically excited by turbulent convection. At least 12
stars have been observed by various groups looking for solar-like
oscillations \citep[cf.\ ][]{BK06}. The amplitudes of these velocity
variations depend on the stellar luminosity and mass and can be scaled
from the Sun, at least for objects that have not evolved too far off
the main sequence \citep{KB95}. These successful detections of
solar-like oscillations have been in no small part due to advances in
precision radial velocity techniques made by planet search teams;
however, little quantitative work has been done to examine the impact
of the oscillations on the detection of extra-solar planets.
                                   
\citet{TBM05} discussed asteroseismology as noise in the
context of Doppler velocity planet searches. They suggested that
exposure times of integer multiples of the peak oscillation periods
could lower jitter by 1-2\,m\,s$^{-1}$. This was also discussed by
\citet{MPQ03}, who argued that exposure times of around 15 minutes
would average out oscillations. To date these suggestions have not
been quantitatively investigated.

\section{Observations}
\label{sec:obs}

To quantify the impact of $p$-mode oscillations on Doppler planet
searches as a function of both observing strategy and stellar
properties we have analysed data from four of the published programs
of high time-resolution observations obtained with the University
College London Echelle Spectrograph (UCLES) that detected and analysed
solar-like oscillations in $\alpha$ Cen A \citep[G2V;][]{BBK04},
$\alpha$ Cen B \citep[K1V;][]{KBB05}, $\beta$ Hyi
\citep[G2IV;][]{BBK01} and $\nu$ Indi \citep[G0IV;][]{BBC06}. We note
that while several of these papers used data obtained with the CORALIE
and UVES spectrographs, we have access only to the UCLES data, so
these other data are not analysed here.

The observations examined in this paper are described in detail in the
references given above and are almost the same as used for the
AAPS \citet{BTM01}. Briefly, the
data were taken using UCLES mounted at the coud\'e focus of the
Anglo-Australian Telescope (AAT). An iodine absorption cell is placed
in the beam, imprinting a forest of molecular iodine absorption lines
onto the stellar spectrum. These lines are used as a wavelength
reference to derive high-precision velocities as described in
\citet{BMW96}. The integration times for each object depend on a
number of factors including its brightness, its expected dominant
oscillation period, $P_{\mathrm{max}}$ and current weather
conditions. The data analysed in this paper is exactly
the same as that used for the asteroseismological analyses described
in the references above, including the removal of long-term drifts in
the velocity time series for all stars except $\nu$ Indi (where such a
correction was not found necessary).

\section{The Effect of Oscillations}
\label{sec:effect}

Observed stellar radial velocity curves show variations from several
sources: those intrinsic to the star; reflex motion due to a
companion; and changes and drifts in the instrumental setup. This
paper investigates the observational effect of the first of these --
in particular, stellar oscillations -- on the second. 

\subsection{Oscillations as noise}
\label{sub:oscnoise}

\begin{figure}
  \begin{center}
    \leavevmode
    \epsfig{file=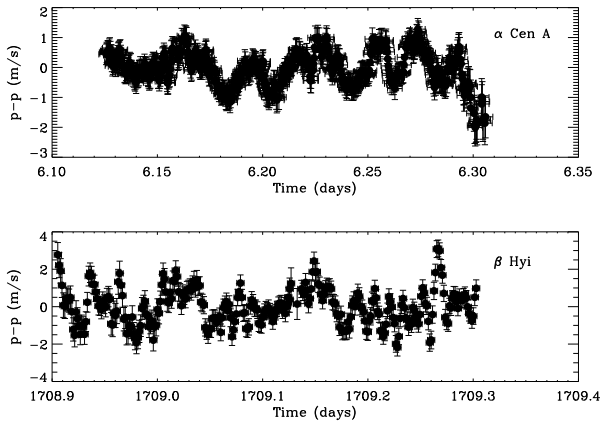,scale=0.47}
    \caption{Moving average with 600\,s window for a set of
      observations of $\alpha$ Cen A and $\beta$ Hyi.}
    \label{fig:movavg}
  \end{center}
\end{figure}

To look at the impact of exposure time on the velocity variations that
$p$-mode oscillations produce in these stars, we have calculated a
moving average of our high time-resolution asteroseismology data
sets. The window of the moving average was set to typical
AAPS exposure times (5, 10, 15, 20,
30, 45, and 60 minutes) as well as several longer times to examine the
effects of averaging an entire night's asteroseismology observations
(90, 120, 180, 300, 450 and 600 minutes). An example of averaged data
over time for $\alpha$ Cen A and $\beta$ Hyi is shown in Figure
\ref{fig:movavg} with a window of 10 minutes. The timestamp is set
to the midtime of the observations in the window. Ten minute exposures
sample almost a cycle and a half of the dominant periodicity in $\alpha$
Cen A, but significantly less than a cycle for $\beta$ Hyi (taking the
dominant period as the individual mode with the highest amplitude). In
both cases, there is still significant scatter in the time series
when the total integration time spans ten minutes.

\begin{figure}
  \begin{center}
    \leavevmode
    \epsfig{file=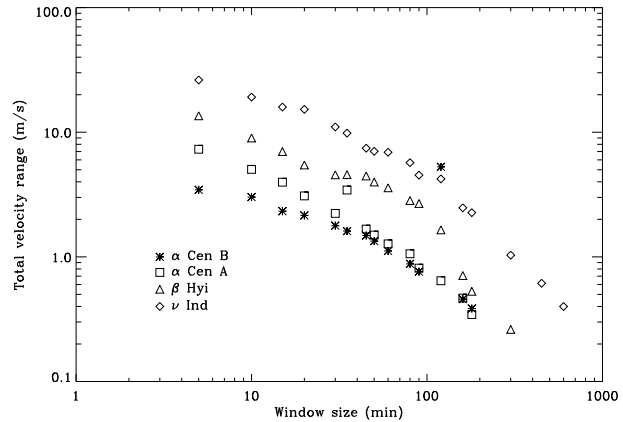,scale=0.47}
    \caption{The total velocity range in an observation data set for
      each star, as a function of the moving average window size.}
    \label{fig:maxmavg}
  \end{center}
\end{figure}

\begin{figure*}
  \begin{center}
    \leavevmode
    \epsfig{file=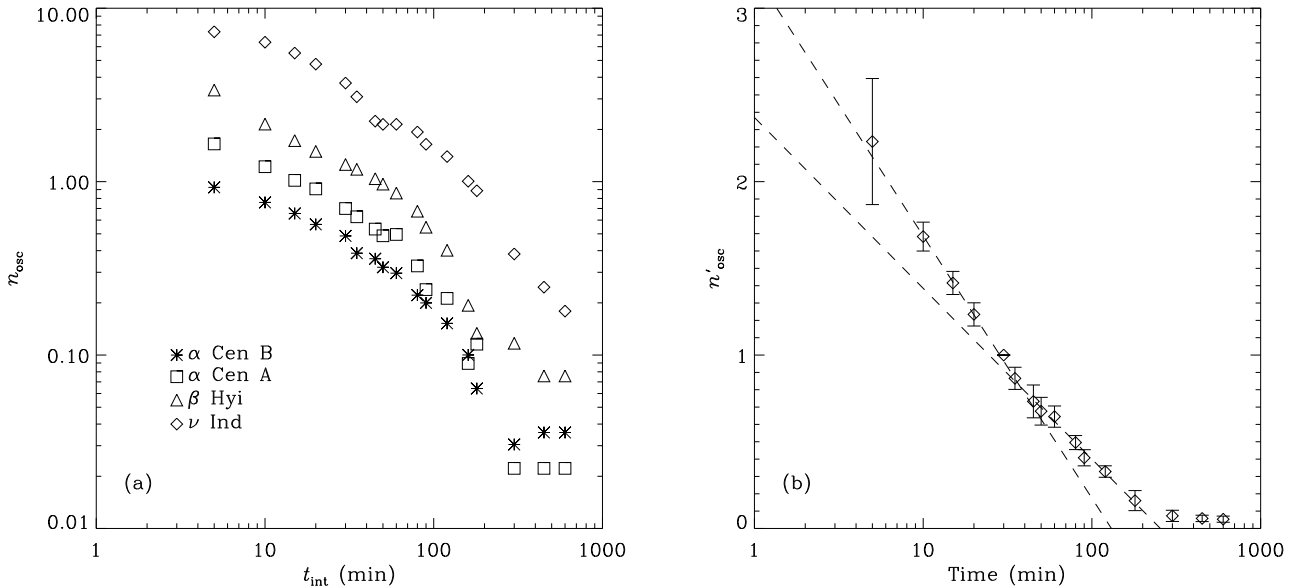,scale=0.70}
    \caption{(a) Ninety-five percent confidence ranges
      ($n_\mathrm{osc}$) for each star as a function of simulated
      integration time ($t_\mathrm{int}$). (b) Data from the panel (a)
      normalised to produce $n^\prime_\mathrm{osc}$ and
      averaged. Uncertainties represent the standard deviation; the
      standard error of the mean is a factor of two smaller.}
    \label{fig:p2p}
  \end{center}
\end{figure*}

If we measure the total velocity range spanned by the velocity
extrema for each star and each windowing time listed above, we produce the
results shown in Figure \ref{fig:maxmavg}. These extrema are measured
from all nights in the time series. If we consider the 10 minute
window for $\alpha$ Cen A shown in Figure \ref{fig:movavg}, we see
that the star varies by up to around 5\,m\,s$^{-1}$. This number
decreases considerably the more cycles we "integrate" over, as
suggested by \citet{TBM05}. 

The size of the variations should vary with spectral type, since
oscillation amplitudes are dependent on stellar luminosity and
mass. Based on equation 7 of \citet{KB95}, earlier-type stars should
be more affected than later-type stars, and subgiants can be
expected to show the largest effects. From Figure \ref{fig:maxmavg} we
see that this is the case and
that $\alpha$ Cen A has higher scatter due to oscillations
than the later-type star $\alpha$ Cen B -- particularly at short total
integration times. Subgiants show larger scatter due to oscillations
than main-sequence stars, with the more evolved subgiant $\nu$ Ind
showing more scatter than the less evolved $\beta$ Hyi. We note that
while the scatter is problematic for planet searches, the oscillations
which cause it allow precise determination of the stellar mass, in
turn leading to more precise planetary masses.

\subsection{Confidence limits and an empirical relation}
\label{sub:empirical}

While the velocity range data of Figure \ref{fig:maxmavg} reveal the
importance of stellar oscillations to precision Doppler planet
searches, what is really required is an understanding of their
statistical impact. That is, how can they be modelled as a source of
Doppler noise for planet searches?

Doppler programs typically model their noise sources as Gaussian
distributions, and in an ideal world, the ``noise impact'' of these
stellar oscillations would be parameterised in the same way. However,
as even a cursory glance at a stellar oscillation power spectrum
indicates \citep[e.g.][]{BBK04}, they are typically \emph{not} a
source of Gaussian noise. Here we are not referring to the underlying
noise of the observations in the absence of oscillations, rather to
the contribution of the oscillations themselves. We have therefore
parameterised the impact of oscillations using ``95\% confidence
velocities'' -- i.e. the velocity range, $n_\mathrm{osc}$, within
which 95\% of the measured velocities for a given target would lie for
a given simulated integration time, $t_\mathrm{int}$.\footnote{Note
that if the asteroseismological power spectrum \emph{were} Gaussian, a
95\% confidence velocity corresponds to a velocity range of
$\pm1.96\,\sigma$ about the mean velocity, or that
$\sigma_\mathrm{osc} \sim n_\mathrm{osc}/4$.}

Figure \ref{fig:maxmavg} suggests there exists a quantifiable
relationship between the jitter due to solar-like oscillations and
total integration time. To examine this Figure \ref{fig:p2p}(a) shows
$n_\mathrm{osc}$ as a function of $t_\mathrm{int}$ for all four stars.
There appears to be a consistent trend in each case, especially at
periods of 5-60\,min. This is not
all that surprising, since while the detailed form of the stellar
oscillation power spectra in these stars \citep[e.g. fine structure
  splitting between modes; ][]{KBB05} will depend on the details of
their interior  structure, the overall envelope of their power
spectra (which is what we sample in these observations) are very
similar. We have normalised each star's $n_\mathrm{osc}$ values at
$t_\mathrm{int}$\,=\,30\,min (which is a typical longest exposure time in the
AAPS for our very highest precision targets) to produce
$n^\prime_\mathrm{osc}$, which we then average over all four stars and
plot in Figure \ref{fig:p2p}(b); the uncertainties are a simple standard
deviation. We have modelled this as two linear trends with a break
point at 35 minutes. We find

\begin{equation}
n^\prime_\mathrm{osc}=\left \{ \begin{array}{lc}
  3.20-1.51\log_{10}t_\mathrm{int} & \mathrm{for}\ t_\mathrm{int}\la 35\,\mathrm{min}\\
  2.37-0.99\log_{10}t_\mathrm{int} &  \mathrm{for}\ 35\la t_\mathrm{int}\la180\,\mathrm{min} \end{array} \right.
\label{eq:scale}
\end{equation}
It is clear from Figure \ref{fig:p2p} that this relationship
breaks down above $t_\mathrm{int}\sim$\,180\,min. This is not
surprising, as almost all of the
power above this time-scale has been extracted by the high-pass filtering of
the asteroseismology data to remove long-term drifts.

\begin{figure}
  \begin{center}
    \leavevmode
    \epsfig{file=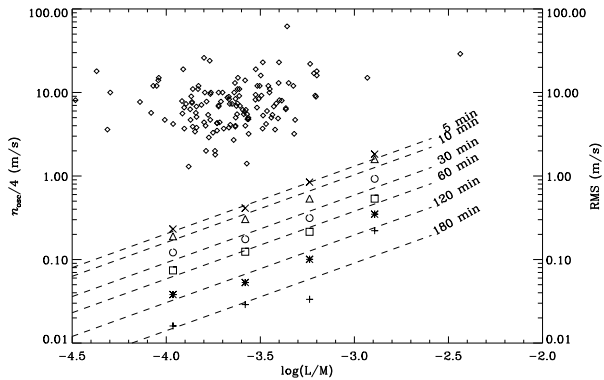,scale=0.42}
    \caption{Oscillation jitter ($n_\mathrm{osc}/4
      \approx\sigma_\mathrm{osc}$) of the UCLES asteroseismology
      targets as a function of $\log_{10}(L/M)$ for various simulated
      integration times: 5 minutes (crosses); 10 min (triangles); 30
      min (circles); 60 min (squares); 120 min (asterisks); 300 min
      (plus signs). Overplotted are the power laws dervied from
      Equation \ref{eq:logp2p}. Finally, the residual RMS values for
      known planets from \citet[][small diamonds]{BWM06} are also
      shown.}
    \label{fig:loglum}
  \end{center}
\end{figure}

We expect ``oscillation noise'' to depend on the ratio of stellar
luminosity to mass in a
similar manner to the oscillations themselves \citep{KB95}. To examine
this, we plot $n_\mathrm{osc}/4$ in Figure \ref{fig:loglum} for each
star as a function of its luminosity-mass ratio ($L/M$), at a range of
simulated exposure times. ($n_\mathrm{osc}/4$ is plotted as this form
is more directly comparable with 1-$\sigma$ noise estimates from other
sources.)

That $n_\mathrm{osc}/4$ has approximately the same slope for
each exposure time on a logarithmic scale suggests a power law
relationship exists with $L/M$. We fit a power law at
$t_\mathrm{int}=30$\,min and find 
\begin{equation}
n_\mathrm{osc}=n^\prime_\mathrm{osc}\times 10^{3.11}(L/M)^{0.92}\ \ \  \mathrm{for}\ t_\mathrm{int}\la180\,\mathrm{min}\\
\label{eq:logp2p}
\end{equation}
where $n^\prime_\mathrm{osc}$ is given by equation
\ref{eq:scale}. The fit is overplotted in Figure \ref{fig:loglum}
along with power laws scaled to the other values of $t_\mathrm{int}$
shown using the appropriate $n^\prime_\mathrm{osc}$.

Also shown in Figure \ref{fig:loglum} are the residual
root-mean-square (RMS) values for each known planet from
\citet{BWM06}. The RMS represents the average
deviation from a perfect fit to the radial velocity data and is made
up of several components, including the various forms of jitter,
instrumental variations and data quality. It is also worth noting here
that low-mass undetected planets are also a source of noise.

The impact of oscillation jitter can be seen (in general) to skirt the
lower edge of the observed exoplanetary RMS values, though for very
short exposure times, or evolved stars, the noise impact reaches
amplitudes of several m\,s$^{-1}$, where it clearly becomes
significant and of concern.

\section{Discussion}
\label{sec:disc}

We have demonstrated that the impact of $p$-mode oscillations on
low-amplitude planet searches can be quantified. Unlike the stellar
activity jitter however, oscillation jitter is dependent on
the length of time spent observing a target at any given epoch, so it
therefore affects the observing strategies for planet hunting.

\subsection{Implications for Observing Strategies}
\label{sub:strategy}

Observing strategies should be tailored to minimise
the impact of oscillations. The key factors to consider are the
target's $L/M$ value and the resulting total integration time needed
to lower $n_\mathrm{osc}$ to a point where it drops below the desired
photon-counting signal-to-noise ratio (SNR) requirement. It is very unlikely
that lengthy asteroseismology campaigns will be staged for even a
small fraction of the stars; however, one can use Equations
\ref{eq:scale} and \ref{eq:logp2p} to optimise integration times. More
evolved stars will have a higher $L/M$ and will be more affected by
oscillation jitter, therefore requiring longer integration times,
regardless of the brightness of the object. \citet{HRQ06} discussed
radial velocity variations in giant stars and suggest that variations
could be even larger than predicted here for these objects.

Consider, for example, the bright, slightly evolved star $\mu$ Ara
(HD\,160691; $V$=5.12; $L=1.75L_\odot$; $M=1.15M_\odot$) which has
been the subject of numerous planet discovery papers, and has been
claimed to host up to 4 planets, the smallest producing velocity
amplitudes as low as 3\,m\,s$^{-1}$. This star is very bright, and so
requires integration times of only a few minutes to reach a SNR sufficient to
achieve $\sim1$\,m\,s$^{-1}$ precision or better with the AAT and
UCLES. At $t_\mathrm{int}=1$ \& 5\,min, we find
$n_\mathrm{osc}/4=0.58, 0.39$\,m\,s$^{-1}$ -- sufficient to make
constraining the innermost planet in this system difficult or
impossible. Observations of $t_\mathrm{int}>15$\,min are required to
reduce $n_\mathrm{osc}/4$ below 0.26\,m\,s$^{-1}$, and of $>60$\,min
to reduce $n_\mathrm{osc}/4$ below 0.11\,m\,s$^{-1}$.  

As seen above, the scale of the effect these oscillations have on
Doppler velocity measurements is smaller than that produced by stellar
activity, but is most significant for giant and sub-giant stars, and
at short integration times (i.e. less than a few minutes). Such short
observations times tend to be used only for very bright
stars. However, as planet searches target lower and lower masses, it
is these very same stars that tend to be targeted for the highest
precision observations. So oscillation noise for these stars can be
important and needs to be accounted for in observing strategies.

\subsection{Avenues of Further Investigation}
\label{sub:convec}

Apart from solar-like oscillations and stellar activity, what other
stellar noise sources remain to be quantified? Two sources of
potential noise are stellar activity cycles and stellar
convection. The activity metric of \citet{Wright05} is useful and has
been widely adopted; however, it does not include a time-varying
component, which is certainly present \citep[e.g.][]{MDJ07}. The
timescales of these variations though are much longer than
oscillations -- the order of years rather than minutes. Incorporating
a time dependence into jitter measurements is worthy of investigation,
especially since stellar activity timescales are similar to Jupiter's
orbital period.

Sun-like stars have large convective cells or granules where material
is dredged up from lower layers and mixed to the surface. The process
involves many random surface motions that even when averaged over time
are almost certainly large enough to affect planet search
measurements. Variations are expected to be around 1-2\,m\,s$^{-1}$
\citep{Dravins99} and may have already been observed in $\mu$ Ara
\citep{BBS05}. Granule lifetimes are typically tens of minutes which is
similar to the exposure times used in planet searches. Despite these
characteristics, the impact of convection has not been quantitatively
investigated.

These two effects will form the focus of our ongoing investigation of
the noise sources limiting precision Doppler planet searches.

\section{Conclusions}
\label{sec:conc}

We have shown that the noise impact of stellar oscillations on
precision Doppler velocities obtained in the search for extra-solar
planets can be significant in some circumstances. We have used
asteroseismological data sets to derive relations which quantify that
impact as a function of integration time, and stellar
luminosity-to-mass ratio. These can be used to improve the quality of
Keplerian fits to planet search data. But most importantly, these
relations can drive observing strategies in the search for the
lowest-mass planets around bright and evolved stars.

\vspace{0.3cm}
We would like to acknowledge the following support: PPARC grant
PP/C000552/1 (HRAJ, CGT, SJOT); and ARC grant DP0774000 (CGT). This
research has made use of the SIMBAD database, operated at CDS,
Strasbourg, France, and the NASA's Astrophysics Data System.

\bibliographystyle{mn2e}
\bibliography{%
mnemonic,%
mnemonic-simple,%
planets,%
planets_stat,%
astero,%
stellar%
}

\end{document}